\documentclass[aps,prl,twocolumn,showpacs]{revtex4}
\usepackage{bm}
\usepackage{graphicx}
\usepackage{amsmath}
\usepackage{eufrak}
\usepackage{color}
\newcommand{\nix}[1]{}
\begin{document}

\title{Giant spin-polarized current in a Dirac fermion system at cyclotron resonance}
\author{P.\,Olbrich,$^1$ C.\,Zoth,$^1$ P.\,Vierling,$^1$ K.-M.\,Dantscher,$^1$
G.V.\,Budkin,$^2$ S.A.\,Tarasenko,$^2$ V.V.\,Bel'kov,$^2$ D.A.\,Kozlov,$^3$  Z.D.\,Kvon,$^3$  
N.N.\,Mikhailov,$^3$ S.A.\,Dvoretsky,$^3$ and S.D.\,Ganichev$^1$}
\affiliation{$^1$Terahertz Center, University of Regensburg, 93040 Regensburg, Germany}
\affiliation{$^2$A.F.\,Ioffe Physical-Technical Institute, Russian
Academy of Sciences, 194021 St.\,Petersburg, Russia}
\affiliation{$^3$Institute of Semiconductor Physics, Novosibirsk, Russia}

\begin{abstract}
We report on the observation of the 
giant 
spin-polarized photocurrent
in HgTe/HgCdTe 
quantum well (QW)
of critical thickness at which a Dirac spectrum emerges.
Exciting  
QW of 6.6~nm width by
terahertz (THz) radiation
and sweeping magnetic field we detected a resonant photocurrent.
Remarkably,
the position of the resonance   
can be tuned from negative (-0.4~T) to positive (up to 1.2~T) magnetic fields by means of optical gating.
The photocurent data, accompanied by 
measurements of  radiation transmission as well as Shubnikov-de Haas and quantum 
Hall effects, give an evidence that the enhancement of the photocurrent is 
caused by cyclotron resonance in a 
Dirac fermion system.
The developed theory
shows that the current is
spin polarized and originates from the spin dependent scattering
of charge carriers heated by the radiation.
\end{abstract}

\maketitle

The electron \textit{dc} transport in 
semiconductor systems with 
massless Dirac fermions has recently moved into the focus 
of modern research yielding challenging fundamental
concepts as well as holding a great potential for applications~\cite{Hasan2010, Moore2010, Zhang2011}.
The linear energy spectrum allows the observation of quantum kinetic effects
and, on the other hand, gives a rise to
a new class of phenomena absent in materials with parabolic dispersion.
The massless Dirac fermions are realized in graphene~\cite{CastroNeto2009}, at surface states of bulk 
topological insulators (TI)~\cite{Volkov1985,Hasan2010, Moore2010, Zhang2011}, in edge channels of
two-dimensional TI~\cite{Koenig2007} as well as
in HgTe/HgCdTe QWs of critical thickness~\cite{Bernevig2006, Buettner2011, Kvon2011}.
In the latter case and TIs, the linear energy spectrum
is formed by strong spin-orbit interaction which locks the orbital motion 
of carriers with their spins.
The interest in Dirac fermions in such materials resulted in theoretical consideration and
observation of such fundamental physical phenomena as the quantum spin Hall effect~\cite{Koenig2007, Kane2005, Bernevig2006, Roth2009,  
Hsieh2008},
quantum Hall effect (QHE) on topological surface states~\cite{Bruene2011}, 
magneto-electric effect~\cite{Qi2008, Essin2009}, 
and quantum interference effects~\cite{Checkelsky2009, Peng2010, Chen2010, Lu2011, Liu2012, Kozlov2012}.
A considerable attention has been also given to the {\it nonlinear} 
high frequency (HF) transport phenomena.
A plethora of such effects has been treated theoretically,
including photogalvanics in TI systems~\cite{Hosur2011, Dora2012, Wu2012, Semenov2012}, and second harmonic
generation (SHG)~\cite{McIver2012-2} as well as radiation-induced QHE~\cite{Kitagawa2011} and
topological states~\cite{Lindner2011}.
While a great number of proposals have been published in the last two years,
the number of experiments on the topic is limited so far by a few publications reporting 
the observation of SHG and photogalvanic effects in 3D TIs induced by near infrared 
radiation~\cite{McIver2012-2,Hsieh2011,McIver2012,Kastl2012}.

Here, we report on the observation of a 
$dc$ current excited by THz radiation in HgTe/HgCdTe QWs
of critical thickness. We show that the current is giantly
enhanced at cyclotron resonance (CR) being a few  orders of magnitude
higher than THz radiation excited photocurrents 
detected in other non-magnetic QW structures. 
Due to the non-equidistant energy
separation of Landau levels in systems with linear electron
dispersion, the CR position
is tuned by the variation of carrier density applying optical
gate. 
The
microscopic origin of the current is discussed in terms of the cyclotron motion, 
spin-dependent scattering and Zeeman splitting.
We show that the current is spin-polarized and its enhancement  comes 
from 
constructively contributing 
three factors: strong spin-orbit coupling, large $g$-factor
in HgTe/HgCdTe QWs, and efficient radiation absorption at CR.

\begin{figure}[t]
\includegraphics[width=\linewidth]{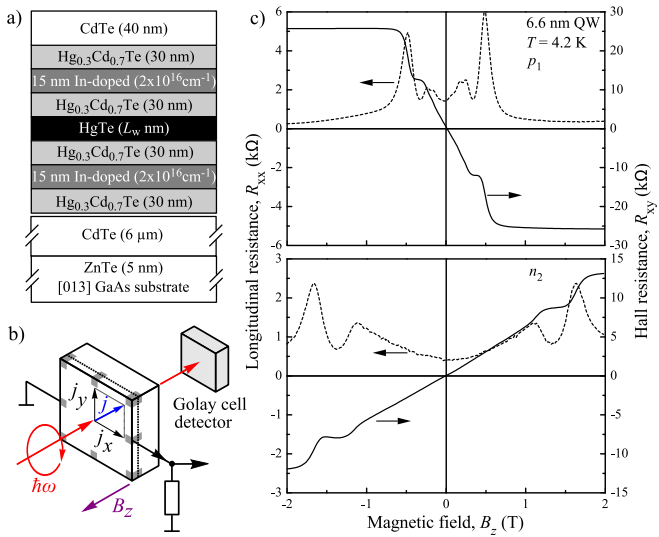}
\caption{ (a) 
Cross section of the investigated
structures. (b) Experimental set up used for the photocurrent and
transmission measurements. (c) Magneto-transport data obtained in the van der Pauw
geometry for 6.6\,nm HgTe/HgCdTe QW sample without (top) and with (bottom) optical
gating. 
The change of slope in the Hall signal indicates
that initially $p$-type sample becomes $n$-type due to optical gating. 
}
\label{fig_transport_6_21nm}
\end{figure}

The experiments are carried out on (013)-oriented
HgTe/Hg$_{0.3}$Cd$_{0.7}$Te QWs~\cite{Kvon2009}.
Single QW samples with
widths, $L_{w}$, of 6.6\,nm, and 21\,nm and mobilities   about
$10^5$\,cm$^2$/{(V$\cdot$s)} at $T = 4.2$\,K are investigated. 
The
structures cross section  
is shown in Fig.~\ref{fig_transport_6_21nm}(a). 
Eight ohmic contacts  
have been prepared at the corners and in the middle of the edges of 
$5 \times 5$\,mm$^2$ 
samples.  
Magneto-transport measurements
show well pronounced Shubnikov-de Haas  
oscillations and QHE plateaus, see
Fig.~\ref{fig_transport_6_21nm}(c). 
To achieve a controllable
variation of the carrier density  we applied
optical gating using the persistent photoconductivity effect well
known for HgTe/HgCdTe QWs~\cite{Kvon2011,Ikonnikov2011,Zholudev2012}. 
We illuminate the sample by red light emitting diode for a time $t_i$
resulting in a change of the carrier density (type) which could be
restored by heating the sample above $T \approx 150$\,K.
The carrier densities measured for different $t_i$ 
used in experiments are given in Table~\ref{table11}.

\begin{table}[b]
\caption{Sample parameters  measured at $T=4.2$~K.}
\begin{tabular}{|c|c|c|c|c|}
\hline
        & \,\, $L_w$, nm & \,\, illum.   & \,\,density, $\rm{cm}^{-2}10^{10}$       & \,\,$E_{\rm F}$, meV         \\
\hline
$p_1$   & 6.6   & -      &  1.5               & 15            \\
\hline
$n_1$   & 6.6   & +      &  3.4               & 21            \\
\hline
$n_2$   & 6.6   & +      & 11.0               & 39            \\
\hline 
\hline
$n_3$   & 21.0  & -      & 18.0               & 15           \\
\hline
$n_4$   & 21.0  & +      & 24.0               & 21           \\
\hline
\end{tabular}
\label{table11}
\end{table}

For photocurrent excitation we apply a $cw$
CH$_3$OH laser emitting a
radiation with frequency $f\,$=$\,2.54$\,THz (wavelength $\lambda\,$=$\,118\,\mu$m)~\cite{Kvon2008}. The radiation with power $P\,$$\approx$$\,10$\,mW  is
focused in a spot of about 1.5\,mm diameter and modulated at 800\,Hz.
Right ($\sigma^+$) and left ($\sigma^-$) handed circularly polarized light is obtained by a $\lambda/4$-plate.
The experimental geometry is sketched in
Fig.~\ref{fig_transport_6_21nm}(b). 
In (013)-oriented QWs, excitation by normally incident THz radiation results in a photogalvanic
current even at $B_z\,$=$\,0$~\cite{Ganichev02,book},
see Supplementary Materials.
Owing low symmetry of QW, the photocurrent has no predefined direction
and its magnitude, $j =\sqrt{j_x^2+j_y^2}$, can be deduced by measuring the
signals along two orthogonal 
directions~\cite{Wittman2010}. The current-induced photovoltages $U_{x,y}$
are picked up across a 1\,M$\Omega$ load resistor applying 
lock-in technique. The magnetic field $B_z$ up to 4\,T is applied normal to the QW plane.
The photocurrent studies are accompanied by optical transmission, see Fig.~\ref{fig_transport_6_21nm}(b),
and magneto-transport  experiments.

\begin{figure}[t]
\includegraphics[width=0.8\linewidth]{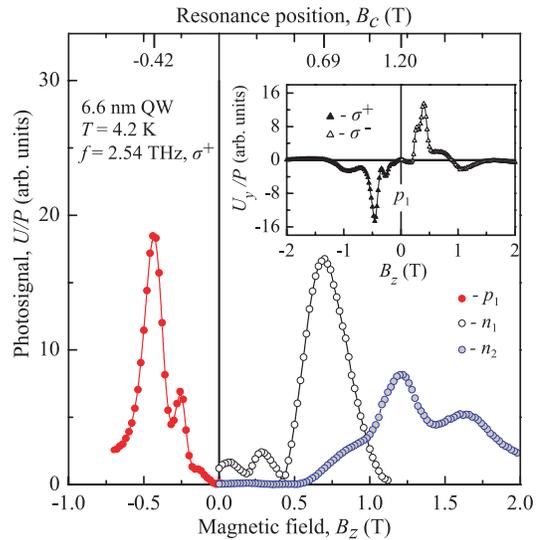}
\caption{ Signal normalized by the radiation power $U/P$ 
excited by
$\sigma^+$
radiation 
vs.
$B_z$. 
The data are shown for magnetic fields lower than the full quantization limit~\cite{footnote1}.
The inset shows data for $\sigma^+$ and $\sigma^-$ light.
 }
\label{fig_circ_PGE_118_4K2_6nm}
\end{figure}

\begin{figure}[t]
\includegraphics[width=\linewidth]{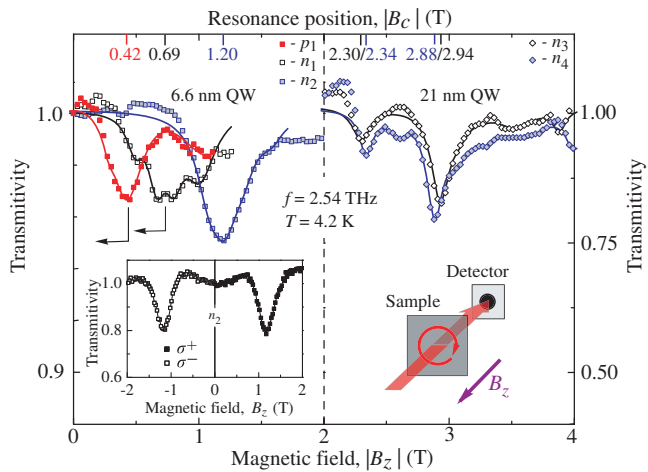}
\caption{Transmissivity of circularly polarized light as a
function of the magnetic field modulus, $|B_z|$.
Full lines are fits
by a superposition
of Lorentz functions.
The inset shows the  
data obtained
for $\sigma^+$
and $\sigma^-$
light for the $L_w\,$=$\,6.6$\,nm  samples.}  
\label{fig_transm_6_21nm}
\end{figure}

We start with the data obtained on the 6.6~nm QW, which should 
have a close to linear dispersion~\cite{Kvon2011,Kozlov2012}.
Exciting the sample with right-handed circularly polarized radiation 
and sweeping  
magnetic field 
we observed  a strong resonant 
photocurrent 
at $B_c = - 0.42$~T, see 
Fig.~\ref{fig_circ_PGE_118_4K2_6nm}. The signal at the resonance
is 
more than two 
orders of magnitude higher than that detected at $B_z\,$=$\,0$. By changing the carrier type from hole
to electron one the resonance jumps from negative to positive $B_z$
and moves towards higher field, being now
for the electron density $n_1\,$$\approx$$\,2 \times p_1$ at $B_c\,$=$\,+ 0.69$\,T.
Remarkably, at further increase in the electron density, the
resonance position drifts to even higher
$B_c$, being 1.2\,T for $n_2\,$$\approx$$\,3 \times n_1$, 
see 
Fig.~\ref{fig_circ_PGE_118_4K2_6nm}. Switching the radiation helicity
from 
$\sigma^+$ 
to 
$\sigma^-$
changes the current sign and mirrors the results with respect to the magnetic field polarity,
as shown for $p$-type conductivity in the inset in Fig.~\ref{fig_circ_PGE_118_4K2_6nm}. The above behavior
is observed for the temperature range from 4.2 up to 150\,K. 
The resonances are detected in the transmission measurements as well. The data for
different carrier densities (type) are shown in
Fig.~\ref{fig_transm_6_21nm} demonstrating a good correlation
between the positions of the dip in the transmissivity and the
resonant photocurrent.

\begin{figure}[t]
\includegraphics[width=\linewidth]{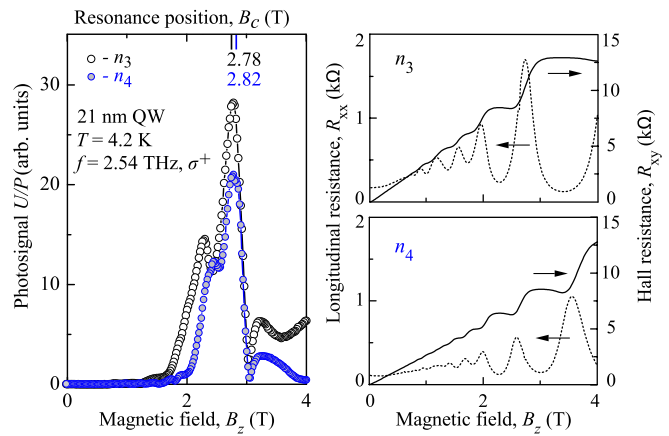}
\caption{Left panel: Magnetic field dependence of the
signal $U/P$ excited by $\sigma^+$ 
light in the 21\,nm sample.
Right panel shows the 
magneto-transport data.
}
\label{fig_circ_PGE_118_184_4K2_21nm}
\end{figure}

Resonant photocurrents are also detected for the 21\,nm QWs, 
a structure characterized by a
nearly parabolic dispersion. Photocurrent (see
Fig.~\ref{fig_circ_PGE_118_184_4K2_21nm})
and 
transmission measurements (see Fig.~\ref{fig_transm_6_21nm})
clearly show that the resonance position in these QWs is shifted to much
higher magnetic fields $B_c\,$$\approx$$\,3$\,T. Furthermore, the resonance field now only
slightly depends on the carrier density

The observed coincidence of peak positions in the
photocurrent and transmissivity unambiguously proves that the
resonant current is caused by CR~\cite{footnote}. This is
also supported by the fact that for a fixed radiation helicity
the resonances in the photocurrent and transmissivity
are detected for one polarity of magnetic field only. 
The striking fact is that, depending on optical gating, the resonance for QWs with $L_w\,$=$\,6.6$\,nm  
is detected  for negative as well as for positive magnetic field  
 and its peak position drastically depends on the Fermi level.
These results are in agreement with Ref.~\cite{Kvon2011}  which concludes that the
energy dispersion in similar structures is close to linear.
Such an electron spectrum lacks the band gap and, therefore,
allows an easy transition from $n$- to $p$-type conductivity,
as proved in transport and CR experiments, see Figs.~\ref{fig_transport_6_21nm}-\ref{fig_transm_6_21nm}.
Furthermore, recent study of gated Hall bar 
6.6\,nm QWs samples prepared from the same batch, 
as samples studied here, manifests a weak localization effects even in 
the vicinity of Dirac point giving an additional proof for existence of gapless 
dispersion in these QWs~\cite{Kozlov2012}.
As a matter of fact, the cyclotron frequency in a system with linear
dispersion depends strongly on the Fermi energy - a characteristic
behavior observed in our 6.6\,nm sample. Indeed, the cyclotron
frequency is described by the well known expression $\omega_c\,$=$\,|e B_z| / m_c c$, where $e$ is the  carrier  charge, 
$c$ is the speed of light, and $m_c$ is the
effective cyclotron mass at the Fermi energy. The latter, given by  $m_c = p_{\rm F}/(d E_F / d p_{\rm F})$
with $p_{\rm F}$ being the Fermi momentum, yields $m_c = E_{\rm F} / v^2$ for a system with linear 
dispersion characterized by a constant velocity $v$. Taking $E_{\rm F}\,$=$\,\sqrt{2\pi n}
(\hbar v$) into account, we obtain for the CR position $|B_c| = \sqrt{2 \pi n} (c
\hbar \omega)/|e v| $. From the resonance positions
measured for electron densities $n_1$ and $n_2$
(Fig.~\ref{fig_circ_PGE_118_4K2_6nm}) we find that the electron
Fermi velocity is almost constant, being equal to $7.2 \times 10^5$~m/s. 
The value is in a good agreement with the electron
velocity  for 2D Dirac fermions in HgTe/HgCdTe QWs of critical
thickness, $v = 6.3 \times 10^5$~m/s, obtained from the energy
spectrum calculated in Ref.~\cite{Buettner2011}. The hole velocity, deduced from our
data for the density $p_1$, is also close to this value ($7.5 \times 10^5$~m/s).
Note that the obtained values are close to the carrier velocity in graphene
($10^6$~m/s~\cite{CastroNeto2009}). The substantially higher
resonance field in 21\,nm QWs as
well as the observed weak dependence of its position on the
electron density
correspond to the CR behavior in HgTe/HgCdTe QWs characterized by a nearly parabolic dispersion
with large Zeeman splitting~\cite{Ikonnikov2011,Zholudev2012,Meyer1992,Truchsess1996,Schultz1998,spirin10}.

\begin{figure}[t]
\includegraphics[width=0.75\linewidth]{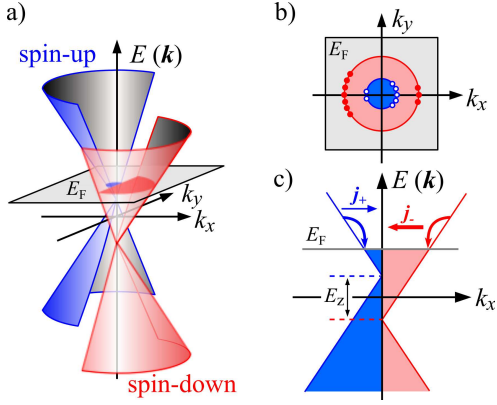}
\caption{Microscopic model.
(a) Energy dispersion 
with spin-up (left, blue) and spin-down (right, red) cones 
shifted due Zeeman effect to  higher and lower energies, respectively. 
(b) Non-equilibrium carrier distribution at Fermi energy caused by the 
spin-dependent scattering. Full and open circles sketch the electron 
distribution in ${\bm k}$-space.
(c) Energy relaxation of an electron gas heated as a result of the CR absorption.
Due to the spin dependent scattering, the relaxation rates for electrons with positive 
and negative ${\bm k}$ are different. Bent arrows show schematically a predominant 
energy relaxation in the spin-up (left) ans spin-down (right) cones. 
The scattering asymmetry within each cone results in the oppositely directed fluxes
$\bm{j}_{+}$ and $\bm{j}_{-}$ shown by horizontal arrows.
As due to the Zeeman effect the spin-down cone  is larger populated, 
the flux $\bm{j}_{-}$  is stronger than $\bm{j}_{+}$ 
and a $dc$  electric current emerges. 
}
\label{model}
\end{figure}

Now we turn to the microscopic origin of the current generation at cyclotron resonance. 
The fact that the current is excited at normal incidence of radiation implies
that it is inherently related to asymmetry of carrier relaxation (excitation) in $\bm{k}$-space
in (013)-oriented QWs of low spatial symmetry~\cite{Ivchenko_book}. In HgTe/HgCdTe QWs with known strong spin-orbit 
coupling and enhanced magnetic properties, the most likely candidate responsible for the microscopic origin 
of the asymmetry is spin-orbit coupling, in particular, spin-dependent scattering. The latter is a known fact for 
other III-V heterostructures~\cite{Nature06,Belkov08}. We consider the
resonant absorption of THz radiation, which leads to the strong electron (hole) gas
heating. The resulting steady-state non-equilibrium
carrier distribution is formed by 
the energy gain due to the radiation absorption, electron-electron collisions 
thermalizing the electron gas, and the energy loss due to emission of phonons.
The matrix element of electron scattering by phonon contains
asymmetric spin-dependent terms (odd in the electron wave vector), which are similar to
the Rashba and Dresselhaus spin-orbit terms in the energy dispersion~\cite{Dyakonov2008}.
Due to the spin-dependent part of
the electron-phonon interaction, the energy relaxation of
carriers in the spin cones is asymmetric and the relaxation
rates for positive and negative wave vectors, say in $x$-direction,
are different~\cite{footnote2}. The asymmetry causes imbalance in
the carrier distribution in {$\bm k$}-space and, hence, 
electron fluxes, see Fig.~\ref{model}(b). The latter have opposite directions
in the spin-up and spin-down cones. 
As, besides the cyclotron motion, the magnetic field  
splits the spin cones due to the Zeeman effect, one of 
the cones is preferentially populated compared to the other. 
Consequently, the fluxes in
the spin cones are unbalanced, and a net electric current emerges.
This process is sketched in Fig.~\ref{model} and, besides the
linear dispersion, is alike that known for other non-magnetic QW
structures~\cite{Nature06,Belkov08}. Obviously, the current
magnitude is proportional to the radiation absorption, the
strength of spin-orbit coupling, and the Zeeman splitting.
The absorption is strongly enhanced in HgTe/HgCdTe QWs at the CR
condition, resulting in the giant photocurrent observed in the
experiments. While the models of the current formation for QWs with linear 
and parabolic dispersions are similar, the position of CR and its behavior
upon variation of the Fermi energy are different. 
In particular, for a fixed radiation frequency, the
cyclotron frequency is almost independent of the carrier density
in QWs with nearly parabolic spectrum. By contrast, for QWs with
linear spectrum, it drastically depends on the carrier density and
for circularly polarized light may even change its sign in the
same structure.

Following the model above, we now develop the quasi-classical theory of the effect. 
In this approach, the electron fluxes in the cones are 
given by $\bm{j}_s = e \sum_{\bm{k}} \bm{v} \, \delta f_{s\bm{k}}$, 
where $\delta f_{s\bm{k}}$ is the anisotropic part of the distribution 
function to be found from the kinetic equation, 
$s$ is the index enumerating the Dirac cones. Here, 
$s=$``$+$''  corresponds to the cone formed from the states
 $|E1,+1/2\rangle = f_1(z)|\Gamma_6,+1/2\rangle + f_4(z)|\Gamma_8,+1/2\rangle$ and 
 $|H1,+3/2\rangle = f_3(z)|\Gamma_8,+3/2\rangle$,  $s=$``$-$'' 
 corresponds to the cone formed from the states 
 $|E1,-1/2\rangle = f_1(z)|\Gamma_6,-1/2\rangle + f_4(z)|\Gamma_8,-1/2\rangle$ and 
 $|H1,-3/2\rangle = f_3(z)|\Gamma_8,-3/2\rangle$, 
 with $f_1(z)$, $f_3(z)$, and $f_4(z)$ being the envelope functions~\cite{Bernevig2006}.

We consider that at low temperatures, relevant to the experimental 
conditions, the momentum relaxation of carriers is limited by elastic 
scattering from static defects while the energy relaxation is governed 
by deformation interaction with bulk acoustic phonons. The spin-dependent 
asymmetry of electron-phonon interaction in (013)-grown QWs is caused by the strain-induced 
coupling between the $\Gamma_6$ and $\Gamma_8$ band 
states~\cite{Pikus88,Ivchenko04,Tarasenko08} and deviation 
of the QW plane from the (001) plane by the
angle $\theta\approx 18.4^{\circ}$.
Then, for the linear electron dispersion in QWs with symmetric confinement potential,
the resulting fluxes have the form (see Supplementary Material)
\begin{equation}\label{j_x}
j_{\pm,x} = \mp \frac{|e|v \sin 2 \theta}{2\sqrt{2}} \frac{n_{\pm}}{n} \left( \frac{d}{d E_F}
 \frac{\omega_c \tau_p^2}{1+(\omega_c\tau_p)^2} \right) \xi I \eta \:,
\end{equation}
\[
j_{\pm,y} = \mp \frac{ev \sin 2 \theta}{2\sqrt{2}} \frac{n_{\pm}}{n} 
\left( \frac{d}{d E_F} \frac{\tau_p}{1+(\omega_c\tau_p)^2} \right) \xi I \eta  \:.
\]
Here $x \| [100]$ and $y \| [03\bar{1}]$ are the in-plane axes,
$n_{\pm}$ are the carrier densities in the cones, 
$n = n_{+} + n_{-}$, 
$\tau_p$ is the momentum relaxation time, $I$ is the radiation intensity, 
$\eta$ is the free-carrier absorbance, $\xi$ is a dimensionless parameter,
\[
\xi = \dfrac{\int_{-\infty}^{+\infty} \Xi_{cv} Z_{13} ( \Xi_c Z_{11} + \Xi_v Z_{33} + \Xi_v Z_{44}) q_z^2 dq_z} 
{\int_{-\infty}^{+\infty} [( \Xi_c Z_{11} + \Xi_v Z_{44})^2 + (\Xi_v Z_{33})^2] q_z^2 dq_z} \:,
\]
$Z_{ij}=\int_{-\infty}^{+\infty} f_i(z) f_j(z) \exp(iq_zz) dz$, 
$\Xi_{c}$ and $\Xi_{v}$ are the deformation-potential constants in the bands $\Gamma_6$ and $\Gamma_8$, 
respectively, and $\Xi_{cv}$ is the interband deformation-potential constant~\cite{Ivchenko04,Tarasenko08}. 
  
For the circularly polarized radiation $\eta$ in the vicinity of CR has the form
\begin{equation}\label{absorbance}
\eta = \frac{2 e^2 E_F}{c \, \hbar^2 n_{\omega}} \frac{\tau_p}{1+(\omega-\omega_c)^2\tau_p^2} \:,
\end{equation}
where $n_{\omega}$ is the refractive index. Assuming that 
the magnetic field splits the electron states due to the Zeeman effect  
but does not affect the scattering and taking into account that in experiment
$\omega_c \tau_p \gg 1$, we estimate the net 
electric current $\bm{j}=\bm{j}_{+}+\bm{j}_{-}$ as
\begin{equation}
j = \frac{c \sin 2\theta}{2\sqrt{2} v} \frac{g \mu_0}{E_F} \xi I \eta \:,
\end{equation}
where
$\mu_0$ is the Bohr magneton. Following the spectral behavior of the absorbance $\eta$
the current $j$ exhibits a sharp peak at CR, whose position depends on the Fermi 
energy, in accordance with the experiment, see Fig.~\ref{fig_circ_PGE_118_4K2_6nm}.

To summarize, we demonstrate that CR absorption by Dirac fermions
in HgTe/HgCdTe QWs of critical thickness 
results in a resonant spin polarized electric current.
The effect is very general and can be observed in other Dirac fermion systems 
with a strong spin orbit coupling, e.g., surface states in 3D topological insulators like
Bi$_2$Se$_3$ and Bi$_2$Te$_3$.
For the latter case such a study is of especial interest, 
because these crystals are centrosymmetric and,  due to symmetry reasons,
the photocurrent emerges only at surface.
Consequently, it provides a unique selective access to fine details of their
band structure like, e.g., effective mass and group velocity, as well as to the spin transport
and spin-dependent scattering anisotropy~\cite{Giglberger2007,Kohda2012}. 
Finally, large resonant currents detected 
at low magnetic fields, about $0.5$\,T for $2.5$\,THz, 
indicate that HgTe/HgCdTe QWs of critical thickness 
are a good candidate for frequency selective CR-assisted detectors
similar to that based on photoconductivity in bulk InSb~\cite{bookTHz},
but operating at about 10 times lower magnetic fields.

\acknowledgments  We acknowledge useful discussions with R. Winkler and E.\,G. Novik.
The  
support from the DFG,
the Linkage Grant of IB of BMBF at DLR, RFBR, RF President grant MD-2062.2012.2, and the ``Dynasty'' Foundation is gratefully acknowledged.

\section{Supplementary Material}

\section{Photogalvanic effects at zero magnetic field}

\begin{figure}[t]  
\includegraphics[width=\linewidth]{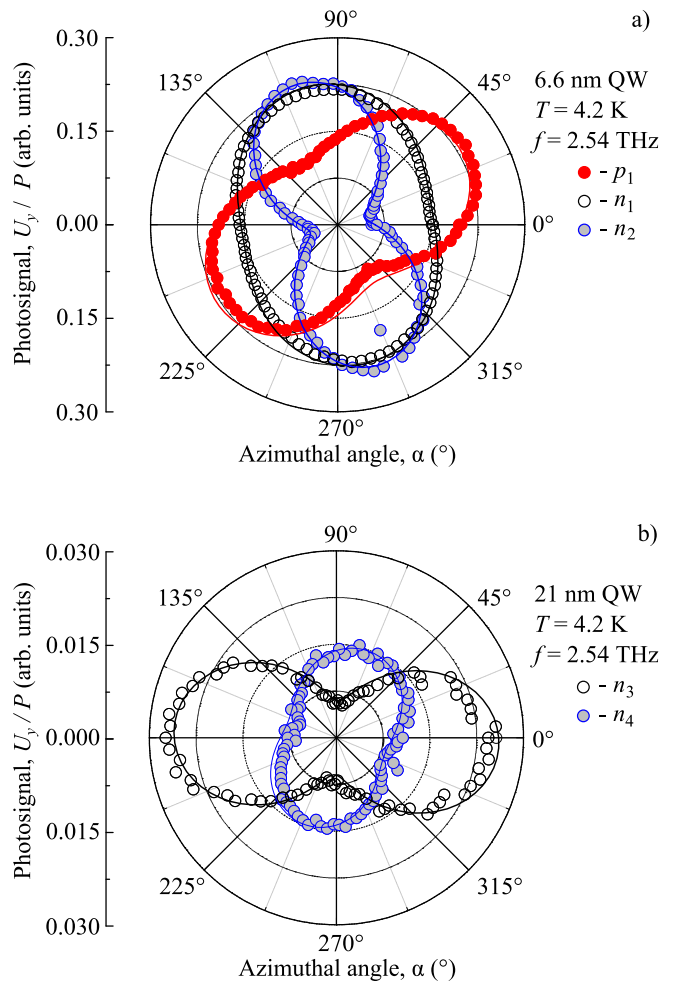} 
\caption{Polarization dependence of the photovoltage normalized by radiation power, $U_y/P$, 
excited with normal incident radiation in (a) $6.6$\,nm and (b) 21\,nm QWs at zero magnetic field.
The normalized photosignal is shown for different carrier 
densities and fitted after $U_y(\alpha)/P = A+ B\sin(2\alpha)+ C \cos(2\alpha)$.
}
\label{Fig_pge_a_polar_118_6nm}
\end{figure}

Illuminating our (013)-oriented QW structures with normally incident 
THz radiation we observed a $dc$ electric current 
even in 
the absence of magnetic field. Figure~\ref{Fig_pge_a_polar_118_6nm} 
shows the  dependence of the
photosignal $U_y/P\,$$\propto$$\,j_y/P$ on the azimuth angle $\alpha$ measured for 6.6 
and 21\,nm QWs, respectively. Here $\alpha$ is the angle between the light polarization plane 
and $x$-direction. In both cases the data are well fitted 
by $U_y(\alpha)/P = A + B\sin(2\alpha) + C \cos(2\alpha)$.
While in 6.6\,nm QWs such a current has not been detected so far, 
in 21\,nm QWs it has been studied in a very details in Ref.~\cite{Wittman2010_sm}
and demonstrated to be due to the photogalvanic effect~\cite{Ganichev02_sm,IvchenkoGanichev_book_sm}. 
A particular feature of the photogalvanic effect in QWs of (013)-orientation orientation 
is that, in contrast to (001)-grown QWs, it can be excited at normal incidence of radiation.
Quantum wells grown on the (013)-oriented substrate belong to the trivial point group C$_1$ lacking any symmetry operation 
except the identity. Hence, symmetry does not impose any restriction on the relation
between radiation electric field and photocurrent components. 
The polarization dependence of the photocurrent in structures of the C$_1$ point-group symmetry 
for the excitation along the QW normal with linearly polarized light
is given by~\cite{Wittman2010_sm}
\begin{eqnarray}
 j_{x} =  \left[ \chi_{xxy} \sin 2\alpha - \frac{\chi_{xxx}+\chi_{xyy}}{2} \right.\\ \nonumber
       \left. +  \frac{\chi_{xxx}-\chi_{xyy}}{2} \cos 2 \alpha \right] I \eta \nonumber
       ,\\
j_{y} = \left[ \chi_{yxy} \sin 2\alpha - \frac{\chi_{yxx}+\chi_{yyy}}{2} \right.\\ \nonumber
\left. +  \frac{\chi_{yxx}-\chi_{yyy}}{2} \cos 2 \alpha \right] I \eta \:, \nonumber
   \label{ph_07_sm}
\end{eqnarray}
where $\bm{\chi}$ is the third rank photogalvanic tensor.
Exactly this polarization dependence 
is observed in experiment, see Fig.~\ref{Fig_pge_a_polar_118_6nm}.
For C$_1$-symmetry group all components of the tensor $\bm{\chi}$ are linearly independent and
may be nonzero. Consequently, even for a fixed light polarization, the CPGE photocurrent direction
is not forced to a certain crystallographic axis. Moreover, it varies with temperature, radiation wavelength, etc~\cite{Wittman2010_sm}.

\section{Microscopic theory}

We describe the electron states in HgTe/CdHgTe quantum wells in the isotropic $\bm{k}$$\cdot$$\bm{p}$ model following Ref.~\cite{Bernevig2006_sm}. The Dirac cones in the QW of critical width are formed from the four states
\begin{eqnarray}\label{basis_states_sm}
|E1,+1/2 \rangle &=& f_1(z) |\Gamma_6,+1/2 \rangle + f_4(z) |\Gamma_8,+1/2 \rangle \:, \nonumber \\
|H1,+3/2 \rangle &=& f_3(z) |\Gamma_8,+3/2 \rangle \:, \nonumber \\
|E1,-1/2 \rangle &=& f_1(z) |\Gamma_6,-1/2 \rangle + f_4(z) |\Gamma_8,-1/2 \rangle \:, \nonumber \\
|H1,-3/2 \rangle &=& f_3(z) |\Gamma_8,-3/2 \rangle \:, 
\end{eqnarray}
which are degenerate at $\bm{k}=0$, with $\bm{k}$ being the in-plane wave vector. Here $f_1(z)$, $f_3(z)$, and $f_4(z)$ are the envelope functions, which can be chosen real, $z$ is the growth direction, $|\Gamma_6,\pm 1/2 \rangle$, $|\Gamma_8,\pm 1/2 \rangle$, and $|\Gamma_8,\pm 3/2 \rangle$ are the basis functions of the $\Gamma_6$ and $\Gamma_8$ bands. At $\bm{k} \neq 0$, the states~(\ref{basis_states_sm}) are coupled that is described by the effective Hamiltonian
\begin{equation}\label{H_eff_sm}
H = \left( 
\begin{array}{cccc}
0 & i A k_+ & 0 & 0 \\
-i A k_- & 0 & 0 & 0 \\
0 & 0 & 0 & - i A k_- \\
0 & 0 & i A k_+ & 0
\end{array}
\right)
\end{equation}
where $A$ is the (real) constant describing the in-plane velocity, $A \approx (P / \sqrt{2}) \int f_1(z) f_3(z) dz$, $P = i (\hbar/m_0) \langle S | p_z | Z \rangle$ is the Kane matrix element, and $k_{\pm} = k_x \pm i k_y$. The sign of $A$ depends on the sign of $f_3(z)$, we take $A>0$. Solution of the Schr\"{o}dinger equation with the Hamiltonian~(\ref{H_eff_sm}) for positive energy $\varepsilon_k = A k$ yields two functions
\begin{equation}\label{functions_sm}
\psi_{+, \bm{k}} = \frac{\exp(i \bm{k} \cdot \bm{\rho})}{\sqrt{2}} 
\left( 
\begin{array}{c}
1 \\
- i \exp(- i \varphi) \\
0 \\
0
\end{array}
\right) \:,
\end{equation}
\[
\psi_{-, \bm{k}} = \frac{\exp(i \bm{k} \cdot \bm{\rho})}{\sqrt{2}} 
\left( 
\begin{array}{c}
0 \\
0 \\
1 \\
i \exp(i \varphi)
\end{array}
\right) \:,
\]
where $\varphi = \arctan (k_y/k_x)$ is the polar angle of the wave vector.

We consider (013)-grown QWs and chose the coordinate frame $x \| [100]$, $y \| [03\bar{1}]$, and $z \| [013]$. In such coordinate system, the basis functions of the $\Gamma_6$ and $\Gamma_8$ bands can be presented in the form
\begin{eqnarray}\label{G6states_sm}
|\Gamma_6,+1/2 \rangle &=& S \uparrow \:, \nonumber \\
|\Gamma_6,-1/2 \rangle &=& S \downarrow \:, 
\end{eqnarray}
\begin{eqnarray}\label{G8states_sm}
|\Gamma_8,+3/2 \rangle &=& - \frac{X' + i (Y' \cos\theta - Z' \sin\theta)}{\sqrt{2}} \uparrow \:, \nonumber \\
|\Gamma_8,+1/2 \rangle &=& \sqrt{\frac23} (Z' \cos\theta + Y' \sin\theta) \uparrow \nonumber \\
&-& \frac{X' + i (Y' \cos\theta - Z' \sin\theta)}{\sqrt{6}} \downarrow \:, \nonumber \\
|\Gamma_8,-1/2 \rangle &=& \sqrt{\frac23} (Z' \cos\theta + Y' \sin\theta) \downarrow \nonumber \\
&+& \frac{X' - i (Y' \cos\theta - Z' \sin\theta)}{\sqrt{6}} \uparrow \:, \nonumber \\
|\Gamma_8,-3/2 \rangle &=& \frac{X' - i (Y' \cos\theta - Z' \sin\theta)}{\sqrt{2}} \downarrow \:,
\end{eqnarray}
where $S$, $X'$, $Y'$, and $Z'$ are the Bloch amplitude of the $\Gamma_6$ and $\Gamma_8$ bands, respectively, referred to the cubic axes $x' \| [100]$, $y' \| [010]$, and $z' \| [001]$, $\theta\approx 18.4^{\circ}$ is the angle between the [001] and [013] axes, and the symbols $\uparrow$ and $\downarrow$ denote the spin projections $+1/2$ and $-1/2$ onto the $z$ axis, respectively. 

The deformation interaction of electrons with acoustic phonons in zinc-blende-type crystals has both intraband and interband contributions~\cite{Pikus88_sm,Ivchenko04_sm}. The matrix elements of strain-induced interband coupling are given by $V_{S,X'} = \Xi_{cv} u_{y'z'}$, $V_{S,Y'} = \Xi_{cv} u_{x'y'}$, $V_{S,Z'} = \Xi_{cv} u_{x'y'}$, where $\Xi_{cv}$ is the interband constant of the deformation potential and $u_{\alpha\beta}$ are the strain-tensor components used here in the primed coordinate system. Note, that $\Xi_{cv} \neq 0$ in non-centrosymmetric crystals only. The matrix elements of strain-induced intraband interaction are taken in the form $V_{S,S}=\Xi_c \, {\rm Tr} \, u_{\alpha\beta}$, $V_{X',X'}=V_{Y',Y'}=V_{Z',Z'}=\Xi_v \, {\rm Tr} \, u_{\alpha\beta}$, where $\Xi_c$ and $\Xi_v$ are the deformation-potential constants ($\Xi_{v}=a$ in the Bir-Pikus notation~\cite{BirPikus_sm}, the constants $b$ and $d$ are neglected for simplicity). Accordingly, the Hamiltonian of electron-phonon interaction in the basis of functions~(\ref{G6states_sm}) and~(\ref{G8states_sm}) has the form
\begin{equation}
V = \begin{pmatrix}
V_c & V_{cv} \\
V_{cv}^{\dag} & V_{v}
\end{pmatrix}\:,
\end{equation}
where $V_c = \Xi_c ({\rm Tr} \, u_{\alpha\beta}) I_{2}$, $V_v = \Xi_v ({\rm Tr} \, u_{\alpha\beta}) I_{4}$,
$I_{2}$ and $I_{4}$ are the identity matrices $2\times2$ and $4\times4$, respectively, 
\begin{widetext}
\[
V_{cv}^{\dag} = \Xi_{cv}
\begin{pmatrix}
- \dfrac{(u_{yz}-iu_{xz}) \cos 2 \theta + (u_{zz}/2-u_{yy}/2+i u_{xy}) \sin 2 \theta} {\sqrt{2}} & \hspace{-5mm} 0 \\
\sqrt{\dfrac23} (u_{xy} \cos 2\theta + u_{xz} \sin 2\theta) & \hspace{-5mm} -\dfrac{(u_{yz}-iu_{xz}) \cos 2 \theta + (u_{zz}/2-u_{yy}/2+i u_{xy}) \sin 2 \theta} {\sqrt{6}} \\
\dfrac{(u_{yz}+iu_{xz}) \cos 2 \theta + (u_{zz}/2-u_{yy}/2-i u_{xy}) \sin 2 \theta} {\sqrt{6}} & \hspace{-5mm} 
\sqrt{\dfrac23} (u_{xy} \cos 2\theta + u_{xz} \sin 2\theta) \\
0 & \hspace{-5mm} \dfrac{(u_{yz}+iu_{xz}) \cos 2 \theta + (u_{zz}/2-u_{yy}/2-i u_{xy}) \sin 2 \theta} {\sqrt{2}}
\end{pmatrix} ,
\]
\end{widetext}
and the strain-tensor components $u_{\alpha\beta}$ are rewritten in the QW coordinate frame.

The dominant contribution to electron scattering is given by the terms proportional to $u_{zz}$
because the out-of-plane component $q_z$ of the wave vector of the phonon involved is typically much larger
than the in-plane component $q_{\|}$. In this approximation, the matrix element of electron scattering from the state $(s,\bm{k})$ to the state $(s,\bm{k}')$, described by the wave functions~(\ref{functions_sm}),
assisted by emission or absorption of a bulk acoustic phonon with the wave vector $\bm{q}$ has the form
\[
V_{s\bm{k}',s\bm{k}}^{(\pm)} = \mp i \frac{q_z}{2} \left[ \frac{\hbar N_{\bm{q}}^{(\pm)}}{2 \rho \Omega_{\bm{q}}} \right]^{1/2} \hspace{-5mm} \left[ \Xi_c Z_{11} + \Xi_v (Z_{44}+ {\rm e}^{i s(\varphi'-\varphi)} Z_{33} )
\right.
\]
\begin{equation}\label{V_kk_sm}
\left. - \frac{i \sin 2 \theta}{2\sqrt{2}} \Xi_{cv}  ( {\rm e}^{i s \varphi'} - {\rm e}^{-i s \varphi}) Z_{13} \right] \delta_{\bm{k}',\bm{k} \mp \bm{q}_{\|}} \:,
\end{equation}
where $N_{\bm{q}}^{(\pm)} = N_{\bm{q}} + (1 \pm 1)/2$, $N_{\bm{q}}$ is the phonon occupation number, $\rho$ is the crystal density, $\Omega_{\bm{q}} = c_l q$ is the phonon frequency, $c_l$ is the speed of sound, $Z_{ij}=\int_{-\infty}^{+\infty} f_i(z) f_j(z) \exp(iq_z z) dz$, and $s=$``$\pm$'' is the index enumerating the Dirac cones. The matrix elements~(\ref{V_kk_sm}) contain asymmetric terms which are responsible for the emergence of oppositely directed electron fluxes $\bm{j}_s$ in the cones during the energy relaxation of heated electron gas. 

To calculate the electron fluxes, we introduce the electron distribution function $f_{s\bm{k}} = \bar{f}_{s k} + \delta f_{s\bm{k}}$, where $\bar{f}_{s k}$ is the quasi-equilibrium function of the Fermi-Dirac type, $\delta f_{s\bm{k}}$ is the anisotropic part of the distribution function. It is assumed that the radiation absorption followed by electron-electron collisions forms the quasi-equilibrium electron distribution with the electron temperature $T_e$ which is slightly higher that the crystal lattice temperature $T_0$. The electron temperature can be found from the energy balance equation
\begin{equation}\label{balance_sm}
\sum_{s,\bm{k},\bm{k}'} W_{s \bm{k}',s \bm{k}}^{({\rm ph})} (\varepsilon_k-\varepsilon_{k'})  \bar{f}_{s k} (1-\bar{f}_{s k'}) = I \eta \:, 
\end{equation}
where $W_{s \bm{k}',s \bm{k}}^{({\rm ph})} = (2\pi/\hbar) \sum_{\bm{q},\pm} |V_{s\bm{k}',s\bm{k}}^{(\pm)}|^2 \delta(\varepsilon_{k'}-\varepsilon_k \pm \hbar\Omega_{\bm{q}})$ is the rate of electron scattering assisted by a phonon emission and absorption, scattering processes between states in different cones are neglected, $I$ is the radiation intensity, and $\eta$ is the free-carrier absorbance. The left-hand side of Eq.~(\ref{balance_sm}) describes the electron energy losses due to cooling by phonons while the right-hand side stands for the energy gain by the free-carrier absorption of radiation.

The electron fluxes are determined by the anisotropic part of the distribution function 
\begin{equation}
\bm{j}_s = e \sum_{\bm{k}} \bm{v} \, \delta f_{s\bm{k}} \:,
\end{equation}
where $\bm{v} = \nabla_{\bm{k}} \varepsilon / \hbar = v \, \bm{k}/k$ is the electron velocity. 
We consider that at low temperatures, relevant to the experimental conditions, the momentum relaxation of carriers is limited by elastic scattering from static defects while the energy relaxation is governed by deformation interaction with bulk acoustic phonons. Accordingly, $\delta f_{s\bm{k}}$ can be found from the Boltzmann equation
\[
\frac{e}{c \hbar} [\bm{v} \times \bm{B}] \cdot \frac{d \, \delta f_{s\bm{k}}}{d \bm{k}} = \sum_{\bm{k}'} [W_{s\bm{k},s\bm{k}'}^{({\rm ph})} \, \bar{f}_{s k'}(1-\bar{f}_{sk})
\]
\begin{equation}
 -
W_{s\bm{k}',s\bm{k}}^{({\rm ph})} \, \bar{f}_{s k}(1-\bar{f}_{sk'})] - \frac{\delta f_{s\bm{k}}}{\tau_p}  \:,
\end{equation}
where $\tau_p$ is the momentum relaxation time. The straightforward calculation shows that the electron fluxes in the cones have the from
\begin{equation}\label{j_x_sm}
j_{\pm, x} = \mp \frac{|e|v \sin 2 \theta}{2\sqrt{2}} \frac{n_{\pm}}{n} \left( \frac{d}{d E_F} \frac{\omega_c \tau_p^2}{1+(\omega_c\tau_p)^2} \right) \xi I \eta \:,
\end{equation}
\begin{equation}\label{j_y_sm}
j_{\pm, y} = \mp \frac{ev \sin 2 \theta}{2\sqrt{2}} \frac{n_{\pm}}{n} \left( \frac{d}{d E_F} \frac{\tau_p}{1+(\omega_c\tau_p)^2} \right) \xi I \eta  \:,
\end{equation}
where $n_{\pm}$ are the carrier densities in the cones, $n = n_{+} + n_{-}$ is the total density, $\omega_c = |e| v^2 B /(c E_F)$ is the cyclotron energy, and    
\[
\xi = \dfrac{\int_{-\infty}^{+\infty} \Xi_{cv} Z_{13} ( \Xi_c Z_{11} + \Xi_v Z_{33} + \Xi_v Z_{44}) q^2 dq} {\int_{-\infty}^{+\infty} [( \Xi_c Z_{11} + \Xi_v Z_{44})^2 + (\Xi_v Z_{33})^2] q^2 dq} \:.
\]
Equations~(\ref{j_x_sm}) and~(\ref{j_y_sm}) are obtained assuming that $E_{F} \gg k_B T \gg \hbar \Omega_{\bm{q}}$ and $|n_{+}-n_{-}| \ll n$.


\begin{thebibliography}{99}
\bibitem{footnote1} Our results reveal that for carrier densities 
$p_1$ [Fig.~\ref{fig_transport_6_21nm}(a)] 
and $n_1$ (not shown) 2D system becomes fully quantized 
at low $B_z$
so that $R_{xy}$ approaches $h/e^2$ 
for $B_z$ above 0.6\,T and 1.2\,T for 
densities $p_1$ and $n_1$, respectively.
\bibitem{footnote2} (013)-grown QWs belong to the $C_1$ point group. Thus, the coupling between the components
of ${\bm k}$ and spin vector 
is not predefined by symmetry.
\end{thebibliography}
\end{document}